\documentclass[twoside,fleqn]{article}
\usepackage{graphicx}
\usepackage{espcrc2}

\title{Model A Dynamics and the Deconfining Phase Transition for
Pure Lattice Gauge Theory\thanks{This 
work was in part supported by the US Department of Energy under 
contract DE-FG02-97ER41022.} }

\author{Alexei Bazavov$\,$\address{ Department of Physics, Florida 
State University, Tallahassee, FL~32306-4350, USA}$^,$\address{ School 
of Computational Science, Florida State University, Tallahassee, 
FL~32306-4120, USA}, Bernd A. Berg$\,^{\rm a, b}$, and Alexander 
Velytsky$\,^{\rm a, b,}$\address{ Present address: Department of 
Physics and Astronomy, UCLA, Los Angeles, CA 90095-1547, USA}
}
\begin{document}
\begin{abstract}
We consider model~A dynamics for a heating quench from the disordered 
(confined) into the ordered (deconfined) phase of SU(3) lattice gauge 
theory. For $4\,N_{\sigma}^3$ lattices the exponential growth factors 
of low-lying structure function modes are calculated.
The linear theory of spinodal decompositions is compared with the data 
from an effective model and the Debye screening mass is estimated from the 
critical mode. 
Further, the quench leads to competing vacuum domains, which make the
equilibration of the QCD vacuum after the heating non-trivial.
We investigate the influence of such domains on the gluonic energy 
density.
\end{abstract}
\date{\today}
\maketitle

\section{Introduction}

In Ref.~\cite{BHMV,BMV} it is argued that a {\it heating} quench of
QCD from its confined phase of disordered Polyakov loops into its 
deconfined phase of ordered Polyakov loops leads to vacuum domains 
of distinct $Z_3$ triality, and one ought to be concerned about 
non-equilibrium effects due to heating of the system. This comes 
because the heating is a rapid quench and the QCD high temperature 
vacuum carries structures which are similar to the low temperature 
phase of analogue spin models. 

Here we report a similar investigation for SU(3) lattice 
gauge theory and preliminary results about the influence of 
such domains on the gluonic energy density and pressure of pure
SU(3) lattice gauge theory. The Markov chain Monte Carlo (MC)
process provides model~A (Glauber) dynamics in the 
classification of Ref.~\cite{ChLu97}. 
As time step a sweep of systematic
updating with the Cabibbo-Marinari \cite{CaMa82} heat-bath
algorithm and its improvements of Ref.~\cite{FaHa84,KePe85} is
used (no over-relaxation, to stay in the universality class of 
Glauber dynamics).
Although this is certainly not the physical dynamics of QCD,
in the present state of affairs it appears important to collect
qualitative ideas about eventual dynamical effects. For this 
purpose the investigation of any dynamics, which actually allows 
for its study ought to be useful.

\section{Preliminaries}

We report numerical results for the structure factors (or 
functions) defined by
\begin{eqnarray} \nonumber
  \hat{S}(\vec{k},t) = 
  \left\langle\hat{u}(\vec{k},t)\,\hat{u}(-\vec{k},t)
      \right\rangle = \\ \label{sf_su3} 
  \int \left\langle u(\vec{r},t)\, u(\vec{r^\prime},t)\ 
  \right\rangle\ \exp\left(i\vec{k}\cdot(\vec{r}
  - \vec{r^\prime})\right)\,d\vec{r}\,d\vec{r'}\,.
\end{eqnarray}
Here $u(\vec{r},t)$ is the relevant fluctuation about some average.
For gauge systems we deal with fluctuations of the Polyakov loop (for 
analogue spin systems with fluctuations of the magnetization). The 
time $t$ corresponds
to the dynamical process, i.e., in our case the Markov chain MC time.
Within the linear theory the differential equation
\begin{eqnarray} \label{u_eq_mot}
  \frac{\partial \hat{u}(\vec{k}^\prime,t)}{\partial t} - 
  \omega(\vec{k}^\prime)\,\hat{u}(\vec{k}^\prime,t)\, =\, g(k^\prime)\\
  \nonumber {\rm with}~~~~ 
  g(k)\, =\, - \Gamma\, f'_0 \sum_{\vec{r}}\exp(i\vec{k}\cdot \vec{r})
\end{eqnarray}
can be derived, where $\Gamma$ is the response coefficient and $f'_0$ 
the derivative of the coarse-grained free energy density with respect 
to the fluctuation variable. The general solution 
of Eq.~(\ref{u_eq_mot}) is 
\begin{eqnarray}
  \hat{u}(\vec{k}^\prime,t) = C\exp\left(\omega(\vec{k}^\prime)\,t\right)
  - \frac{g(\vec{k}^\prime)}{\omega(\vec{k}^\prime)}\\ \label{omega_k}
  {\rm with}~~~~ \omega(\vec{k}) = - \Gamma\, \left( 
  K\,\vec{k}^{\,2} + f^{''}_0 \right)\ .
\end{eqnarray}
Where $K$ is the lowest order expansion coefficient of the free
energy density about its minimum $f_0$, i.e. $K>0$. If $f^{''}_0>0$
holds, Eq.~(\ref{omega_k}) implies that the amplitude of any fluctuation 
approaches a constant exponentially fast with time. But if the second 
derivative is negative, then one sees an exponential growth of the 
fluctuations for momentum modes smaller than the critical value 
\begin{equation} \label{k_c}
  |\vec{k}|<k_c=|\vec{k}_c|=\left[-\frac{f^{''}_0}{K}\right]^{1/2}\,. 
\end{equation}
The equation of motion for the structure factor (\ref{sf_su3}) is 
derived by taking the time derivative of $\hat{S}(\vec{k},t)$ and 
using Eq.~(\ref{u_eq_mot})
\begin{eqnarray} \nonumber
  \frac{\partial \hat{S}(\vec{k},t)}{\partial t} &=& \left\langle 
  2\,\hat{u}(\vec{k},t)\,\hat{u}^*(\vec{k},t)\,\omega(k)\right\rangle \\
  &+& \left\langle \left(\hat{u}^*(\vec{k},t)+\hat{u}(\vec{k},t)
      \right) g(k) \right\rangle\,. \label{s_eq_mot1}
\end{eqnarray}
The average of fluctuations about the mean of the fluctuation variable 
has to be zero $\left\langle\hat{u}(\vec{k},t)\right\rangle=0$. Thus 
(\ref{s_eq_mot1}) becomes
\begin{equation}
  \frac{\partial \hat{S}(\vec{k},t)}{\partial t}
  = 2\,\omega(\vec{k})\,\hat{S}(\vec{k},t)\,,
\end{equation}
with the solution
\begin{equation} \label{str_fact}
  \hat{S}(\vec{k},t)=
  \hat{S}(\vec{k},t=0)\exp\left(2\omega(\vec{k})t\right)\,.
\end{equation}
Again, for  $f^{''}_0<0$ low momentum modes grow exponentially. 
The value of the critical momentum is the same as for the fluctuations. 
Originally the linear theory was developed for model~B 
\cite{CaHi58,Ca68}. Details for model~A can be found in 
Ref.~\cite{BMV}.

During our simulations the structure functions are averaged over 
rotationally equivalent momenta and the notation $S_{n_i}$ is used
to label structure functions of momentum 
\begin{equation} \label{momenta}
  \vec{k} = {2\pi\over L}\,\vec{n}~~~{\rm where}~~~|\vec{n}|=n_i\,.
\end{equation}
We recorded the modes (including the permutations)
$n_1$: $(1,0,0)$, 
$n_2$: $(1,1,0)$, 
$n_3$: $(1,1,1)$, 
$n_4$: $(2,0,0)$,
$n_5$: $(2,1,0)$,
$n_6$: $(2,1,1)$,
$n_7$: $(2,2,0)$,
$n_8$: $(2,2,1)$ and $(3,0,0)$,
$n_9$: $(3,1,0)$, 
$n_{10}$: $(3,1,1)$,
$n_{11}$: $(2,2,2)$,
$n_{12}$: $(3,2,0)$, 
$n_{13}$: $(3,2,1)$,
$n_{14}$: $(3,2,2)$,
$n_{15}$: $(3,3,0)$,
$n_{16}$: $(3,3,1)$,
$n_{17}$: $(3,3,2)$,
$n_{18}$: $(3,3,3)$.
Note that there is an accidental degeneracy in length for $n_8$.

\section{Numerical Results}

\begin{figure}[ht] \vspace{-8mm} \begin{center}
 \includegraphics[width=\columnwidth]{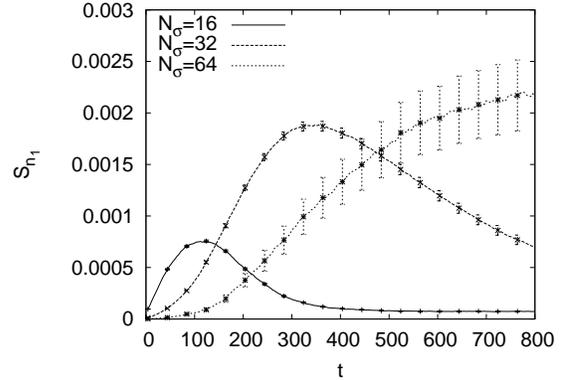} \vspace{-15mm}
 \caption{The first structure function mode for pure SU(3) lattice 
  gauge theory on $4\,N_{\sigma}^3$ lattices. }
 \label{fig_sf01} \end{center} \vspace{-8mm}
\end{figure}

In Fig.~\ref{fig_sf01} the time evolution of the first structure 
function mode after a heating quench from $\beta = 5.5\to 5.92$ in 
pure SU(3) lattice gauge theory is depicted. For the $4\times 16^3$ 
lattice the pseudo-transition 
value is $\beta_c = 5.6908\,(2)$ \cite{BoEn96} and for the larger
lattices the value is expected to be close-by. Our results are
averages over 10,000 repetitions of the quench for the $4\times 16^3$ 
lattice and 4,000 repetitions for the $4\times 32^3$ lattice. 
The $4\times 64^3$ lattices are still running: Presently there are 
only 60 repetitions and only some structure function maxima are presently
reached. Notable in Fig.~\ref{fig_sf01} is the strong increase of the 
maxima $S_{n_1}^{\max}$ with lattice size. In our normalization 
non-critical behavior corresponds to a fall-off $\sim 1/N_{\sigma}^3$ 
and a second order phase transition to a slower fall-off 
$\sim 1/N_{\sigma}^x$ with $0<x<3$. As the $N_{\sigma}\to\infty$ limit 
is bounded by a constant, our figure shows that with our lattice sizes 
the asymptotic behavior has not yet been reached.
	
\begin{figure}[ht] \vspace{-8mm} \begin{center}
 \includegraphics[width=\columnwidth]{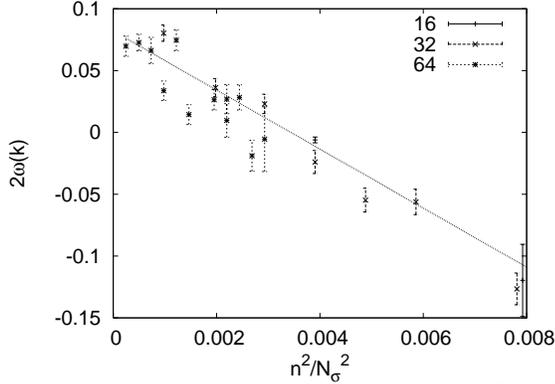} \vspace{-15mm}
 \caption{Determination of $k_c$ for the pure SU(3) lattice gauge
 theory on $4\,N_{\sigma}^3$ lattices. }
 \label{fig_su3_omega} \end{center} \vspace{-8mm}
\end{figure}

To determine the critical mode $k_c$ we fit the initial increase of the 
structure functions. The results are combined in Fig.~\ref{fig_su3_omega} 
and indicate $a\,k_c=2\pi\,n_c/N_{\sigma}\approx 0.34$ (from the figure 
$n_c^2/N_{\sigma}^2\approx 0.003$). Using results of Miller and 
Ogilvie~\cite{MiOg02}, we have $m_D=\sqrt{3}\,k_c $, where $m_D$ is 
the Debye screening mass at the temperature in question. The relation 
$k_c/T_f = N_{\tau}\, a\, k_c$, where $T_f$ is the final temperature 
after the quench, allows us to convert to physical units. For our 
quench we have $T_f/T_c=1.57$ and get 
\begin{equation} \label{Debye}
m_D = \sqrt{3}\, N_{\tau}\, a\,k_c\,T_f = 3.7\, T_c\ .
\end{equation}
For pure SU(3) lattice gauge theory $T_c = 265\,(1)\,$MeV holds,
assuming $\sigma=420\,$MeV for the string tension, while for QCD 
the cross-over temperature appears to be around $T_c\approx 165\,$MeV, 
see \cite{Pe04} for a recent review.

In Fig.~\ref{fig_sf01} we observe that not only the height of the
peaks increases with lattice size, but also the time $t_{\max}$,
$S^{\max}_{n_1}=S_{n_1}(t_{\max})$, which it takes to reach them.
Whereas $S^{\max}_{n_1}$ has finally to approach a constant value,
$t_{\max}$ is expected to diverge with lattice size due to the 
competition of vacuum domain of distinct $Z_3$ triality. 

\begin{figure}[ht] \vspace{-8mm} \begin{center}
\includegraphics[width=\columnwidth]{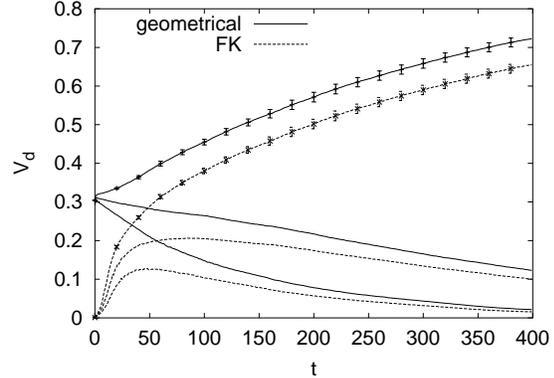} \vspace{-15mm}
\caption{Largest geometrical and FK clusters for the 3D 3-state Potts
model quenched from $\beta=0.2$ to $\beta_f=0.3$ at zero field on a
$40^3$ lattice.} \label{fig_qnch_3d_geom3d} \end{center} \vspace{-8mm}
\end{figure}

Using Fortuin-Kasteleyn (FK) clusters, the competition of such distinct 
vacuum domains can be made visible for the analogue Potts 
models~\cite{BMV}. The states of the three-dimensional, three-state
Potts model substitute then for the $Z_3$ trialities of SU(3) lattice
gauge theory. In Fig.~\ref{fig_qnch_3d_geom3d} we compare the evolution 
of geometrical and FK clusters for a quench of this model from its 
disordered into its ordered phase. We plot the evolution of the largest 
clusters for the three Potts magnetizations in zero external magnetic 
field $h$. While the the system grows competing FK clusters of each 
magnetization before one becomes dominant, geometrical clusters do 
not compete. This picture is unfavorable for the use of geometrical 
clusters of Polyakov loops in gauge theories, for which the FK 
definition does not exist. 

The process of competitions between the largest FK  clusters of different 
magnetization leads for the proper transition ($h=0$) to a divergence
of the equilibration time in the limit of infinite systems, an effect
known in condensed matter physics~\cite{ChLu97}. Potts studies \cite{BMV} 
with an external magnetic field show that a major slowing down effect 
survives for $h\ne 0$. As the influence of an external magnetic field on 
the Potts model is similar to that of quarks on SU(3) gauge theory this
indicates that the effect may be of relevance for QCD studies of the 
crossover region.

For gauge theories a satisfactory cluster definition is presently not 
available.  Nevertheless the underlying mechanism is expected to be the 
same as in the spin models. To study its influence on the gluonic 
energy $\epsilon$ and pressure $p$ densities, we like to calculate
these quantities at times $t\le t_{\max}$. 

Let us first summarize the equilibrium procedure of 
Ref.~\cite{BoEn96,EnKa00} (in earlier work \cite{De89,EnFi90} the 
pressure 
exhibited a non-physical behavior after the deconfining transition and 
the energy density approached the ideal gas limit too quickly due to 
the fact that anisotropy coefficients were calculated perturbatively). 
We denote expectation values of spacelike plaquettes by $P_\sigma$ and 
those involving one time link by $P_\tau$. The energy density and 
pressure can then be cast into the form
$(\epsilon+p)/T^4 =$
\begin{equation}\label{e:eplusp}
  8N_cN_\tau^4g^{-2}\left[1-\frac{g^2}{2}
  [c_\sigma(a)-c_\tau(a)]\right] (P_\sigma-P_\tau)
\end{equation}
and $(\epsilon-3p)/T^4 =$
\begin{equation}\label{e:eminus3p}
  12N_cN_\tau^4\, [ c_\sigma(a)-c_\tau(a) ]
  \left[2P_0-(P_\sigma+P_\tau)\right], 
\end{equation}
where $P_0$ is the plaquette expectation value on a symmetric 
($T=0$) lattice and \textit{anisotropy coefficients}
$c_{\sigma,\tau}(a)$ are defined as follows:
\begin{equation}\label{e:cst}
    c_{\sigma,\tau}(a)\equiv
    \left(\frac{\partial g^{-2}_{\sigma,\tau}}
    {\partial \xi}\right)_{\xi=1}.
\end{equation}
They are related to the QCD $\beta$-function in a simple way
\begin{equation}\label{e:bfun}
    a\frac{dg^{-2}}{da}=-2(c_\sigma(a)+c_\tau(a)).
\end{equation}
One needs to calculate the $\beta$-function and anisotropy coefficients 
non-per\-tur\-ba\-tive\-ly to obtain meaningful results.  In the range 
above the phase transition this is carried out in \cite{BoEn96}. Using 
integral methods and Pade fits from \cite{EnKa00} one finds the spatial 
anisotropy coefficient
\begin{equation}\label{e:csa}
   c_\sigma(a)=c_\sigma(0)\frac{1+d_1g^2+d_2g^4}{1+d_0g^2}
\end{equation}
with $d_0=-0.64907$, $d_1=-0.61630$ and $d_2=0.16965$, where from 
perturbative calculations $c_\sigma(0)=0.150702\,N_c-0.146711/N_c$.
Thus, taking data for the $\beta$-function from Table~3 of
\cite{BoEn96} and evaluating $c_\sigma(a)$ and $c_\tau(a)$
from Eqs. (\ref{e:csa}) and (\ref{e:bfun}) we can
calculate the energy density $\epsilon$ and pressure $p$ using
Eqs. (\ref{e:eplusp}) and (\ref{e:eminus3p}) for arbitrary values
of the coupling $\beta$ above $\beta_c$.
	
\begin{figure}[ht] \vspace{-8mm} \begin{center}
  \includegraphics[width=\columnwidth]{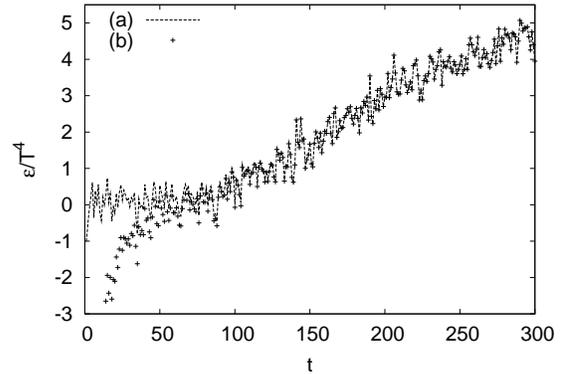} \vspace{-15mm}
  \caption{SU(3) gluonic energy density: (a) with $P_0$ calculated from 
  the time series after the quench and (b) using the equilibrium value 
  of $P_0$.} \label{fig_su3subtract} 
\end{center} \vspace{-8mm} \end{figure}

To normalize to zero temperature, plaquette values from the symmetric 
$N_{\tau}=N_{\sigma}$ lattice are needed in Eq.~(\ref{e:eminus3p}). For 
the quench this leaves us with the question whether we should perform 
the quench also on the symmetric lattice and use $P_0$ from the 
corresponding time evolution, or whether we should take the equilibrium 
value at $\beta=5.92$ for $P_0$. Fortunately, the empirical answer is 
that it does not matter. When one is far enough into the time evolution 
the results agree, see Fig.~\ref{fig_su3subtract} for the time evolution
on our $N_{\sigma}=16$ lattice. The reason is that one stays within the 
confined phase on the symmetric lattice. Therefore its equilibration 
after the quench is fast.
	
\begin{figure}[ht] \vspace{-8mm} \begin{center}
 \includegraphics[width=\columnwidth]{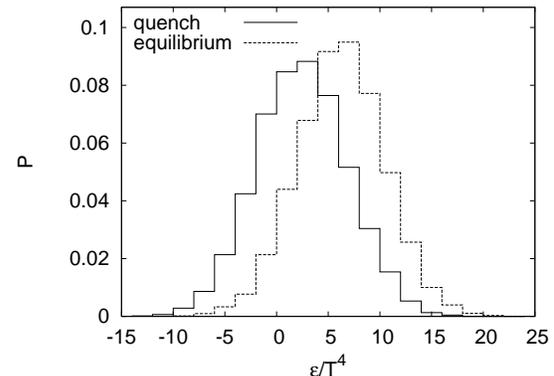} \vspace{-15mm}
 \caption{SU(3) gluonic energy density $P(\epsilon)$ histograms: 
 (a) with competing vacuum domains present and (b) after reaching 
  equilibrium.} \label{fig_su3ged_2hists} 
\end{center} \vspace{-8mm} \end{figure}

Finally, we compare in Fig.~\ref{fig_su3ged_2hists} for the 
$N_{\sigma}=16$ lattice the gluonic energy distribution in equilibrium 
at $\beta=5.92$ with the one obtained after 148 time steps. We find
a shift towards lower gluonic energies and the width of the probability 
density is slightly broader for the time evolution after the quench 
than in equilibrium. 
One also has to take into account that the geometry of relativistic 
heavy ion experiments experiments is reasonably approximated by
$N_{\tau}/N_{\sigma}={\rm const}$, $N_{\sigma}\to\infty$, rather
than by $N_{\tau}={\rm const}$, $N_{\sigma}\to\infty$.

\section{Summary and Conclusions}

Structure functions allow to identify the transition scenario. For our
quench from the disordered into the ordered phase of SU(3) lattice
gauge theory we find spinodal decomposition. Relying on the linear 
theory of spinodal decomposition, we have calculated the critical mode 
$k_c$. From it the Debye screening mass $m_D$ at temperature $T$ is 
determined using  phenomenological arguments of Miller and 
Ogilvie~\cite{MiOg02}. 

With increasing lattice size $N_{\sigma}$ the time to reach the 
structure function maxima diverges. Relying on a study of 
Fortuin-Kasteleyn clusters in Potts models \cite{BMV}, we assume 
that the reason is that vacuum domains of distinct $Z_3$ trialities
compete. These could be the relevant configurations after the heating 
quench in relativistic heavy ion experiments. We have initiated a study 
of the gluonic energy and pressure densities on such configurations.

All our results rely on using a dissipative, non-relativistic
time evolution, believed to be in the Glauber universality class. 
The hope is that the thus created non-equilibrium configurations may 
exhibit some features, which are in any dynamics typical for the 
state of the system after the quench. This hope could get more credible 
by studying a Minkowskian time evolution of Polyakov loops and finding 
similar features. Such a study appears to be possible \cite{Du01} within 
a relativistic Polyakov loop model which was introduced by 
Pisarski \cite{DuPi01}.

\end{document}